\documentclass{ws-xxx}

\newcommand{\ba}{\begin{eqnarray}}
\newcommand{\ea}{\end{eqnarray}}
\newcommand{\be}{\begin{equation}}
\newcommand{\ee}{\end{equation}}
\begin{document}
\title{Ultrahigh Energy Tau Neutrinos}

\author{J. JONES, I. MOCIOIU, I. SARCEVIC} 

\address{Department of Physics \\
University of Arizona\\ 
Tucson AZ 85721\\
}

\author{M.~H.~Reno}
\address{
 Department of Physics and Astronomy\\ University of Iowa\\
Iowa City IA 52242}
\maketitle

\abstracts{
We study ultrahigh energy astrophysical neutrinos and the contribution of
tau neutrinos from neutrino oscillations, 
 relative to the contribution of the other flavors.
We show the effect of tau neutrino regeneration and tau energy loss
 as they propagate through the Earth.   We consider a variety of
neutrino fluxes, such as cosmogenic neutrinos and neutrinos that originate
in Active Galactic Nuclei.
 We discuss signals of tau neutrinos in
detectors such as IceCube, RICE and ANITA.
}

\section{Introduction}

Ultrahigh energy astrophysical neutrinos provide unique probes of particle physics at energies
currently not accessible to colldier experiments.  In addition, neutrinos point
back to their sources and they travel very large distances without interactions,
providing valuable information about extreme enviroments.

The experimental data on $\nu_\mu\leftrightarrow\nu_\tau$ neutrino
oscillations \cite{2} implies that astrophysical sources of muon neutrinos
become sources of $\nu_\mu$ and $\nu_\tau$ in equal proportions after
oscillations over astronomical distances \cite{3}.  Even though $\nu_\mu$ and
$\nu_\tau$ have identical interaction cross sections
at high energies, signals from
$\nu_\tau\rightarrow \tau$ conversions have the potential to contribute
differently from $\nu_\mu$ signals. The $\tau$ lepton can decay
far from the detector, regenerating $\nu_\tau$ \cite{4}.
In case of $\nu_\mu$, muons produced via charged-current decay but 
electromagnetic energy loss coupled with the long
muon lifetime make the $\nu_\mu$ regeneration from muon decays irrelevant
for high energies. Another signal of $\nu_\tau\rightarrow\tau$ is
the tau decay itself \cite{5}.

We have studied in detail the propagation of all flavors of 
neutrinos with very high energy ($E \geq 10^6$ GeV) as 
they traverse the Earth.  Because of the high energies attenuation shadows 
most of the upward-going solid angle at high energies, so we have limited
our consideration to nadir angles larger than $80^\circ$.
We have focused on 
the contribution from tau neutrinos, 
produced in oscillations of extragalactic muon 
neutrinos as they travel large astrophysical distances.  

Neutrinos from astrophysical sources are usually produced via pion
decays, 
which determine the flavor ratio $\nu_e:\nu_\mu:\nu_\tau$ to be
$1:2:0$. 
After propagation over 
very long distances, neutrino oscillations change this ratio to
$1:1:1$ 
because of the 
maximal $\nu_\mu\leftrightarrow\nu_\tau$ mixing. 
For the GZK flux, $\nu_e$ and $\nu_\mu$ incident
fluxes are different because of the additional contributions from 
$\bar{\nu}_e$ from neutron decay and $\nu_e$ from $\mu^+$ decays
\cite{GZK}. 
Because of 
this, the flavor ratio at Earth is affected by the full three flavor
mixing 
and is 
different from $1:1:1$. Given fluxes at the source 
$F^0_{\nu_e}$, 
$F^0_{\nu_\mu}$ and $F^0_{\nu_\tau}$, the fluxes at Earth become:
\ba
F_{\nu_e}&=&F^0_{\nu_e}-\frac{1}{4}\sin^22\theta_{12} (2 F^0_{\nu_e}-F^0_{\nu_\mu}-
F^0_{\nu_\tau})
\label{fle}
\\
F_{\nu_\mu}&=&F_{\nu_\tau}=\frac{1}{2}(F^0_{\nu_\mu}+F^0_{\nu_\tau})+\frac{1}{8}\sin^22\theta_{12}
(2 F^0_{\nu_e}-F^0_{\nu_\mu}-F^0_{\nu_\tau})
\label{flmutau}
\ea
where $\theta_{12}$ is the mixing angle relevant for solar neutrino oscillations. We 
have assumed that $\theta_{23}$, the mixing angle relevant for atmospheric neutrino 
oscillations, is maximal and $\theta_{13}$ is very small, as shown by reactor experiments,
as well as atmospheric and solar data. 

For GZK 
neutrinos\cite{GZK} produced by cosmic ray interactions
with the microwave background, the flavor ratio at 
Earth deviates from $1:1:1$ because the initial fluxes are somewhat different, 
and they start out in a ratio not equal to $1:2:0$. In this case
the $\nu_e\leftrightarrow \nu_\mu,\nu_\tau$ oscillations relevant to 
solar neutrino
oscillations start playing a role, in addition to the maximal 
$\nu_\mu\leftrightarrow\nu_\tau$ oscillations relevant to atmospheric 
neutrinos. 
Z burst neutrinos\cite{zburst} from models with ultrahigh energy
neutrinos scattering with relic neutrinos to produce Z bosons are 
also considered below, where neutrinmixing yields flux ratios
of $1:1:1$.

\begin{figure}[t]
\centerline{\epsfxsize=3.5in\epsfbox{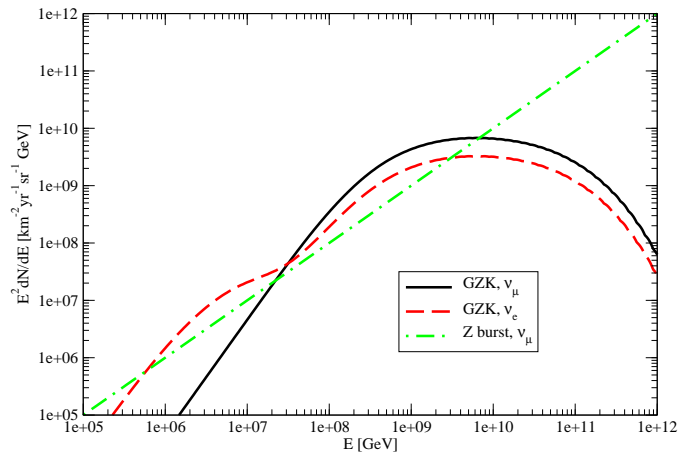}}   
\caption{Initial Neutrino Fluxes}
\label{fluxes}
\end{figure}

The initial fluxes for GZK and Z burst neutrinos are shown in Fig. \ref{fluxes}.
In our propagation of neutrinos and charged leptons through the Earth\cite{1}, 
we have focused on kilometer size neutrino detectors, such as 
ICECUBE~\cite{icecube} and the Radio Ice Cerenkov
Experiment (RICE)\cite{rice} and 
on a detector with much larger effective area which uses Antarctic 
ice as a converter, the Antarctic Impulsive Transient Antenna
(ANITA)\cite{anita}.  

Signals of neutrino interactions in the rock below the ice or in ice
depend on the energy and flavor of the neutrino.
 Muon neutrino charged current (CC) conversions
to muons are noted by the Cherenkov signal of upward going muons in a
detector such as IceCube \cite{icecube}. High energy electromagnetic showers
from $\nu_e\to e$ CC interactions produce Cherenkov radiation
which is coherent for radio wavelengths.  
The Radio Ice Cherenkov Experiment (RICE) has put limits on
incident isotropic electron neutrino fluxes which produce downward-going
electromagnetic showers \cite{rice}. The Antarctic Impulsive
Transient Antenna (ANITA) also uses the ice as a neutrino converter
\cite{anita}. These balloon missions will monitor the ice sheet
for refracted radio frequency signals with an effective telescope area of
1M km$^2$. All flavors of neutrinos produce hadronic showers. In addition,
tau decays contribute to both electromagnetic and hadronic showers that
could be detected by IceCube, RICE or ANITA.

\section{Neutrino Propagation}

Propagation of neutrinos and charged leptons is 
governed by the following transport equations:
$$
\frac{\partial F_{\nu_{\tau}}(E,X)}{\partial X}\!\! 
=-N_A\sigma^{tot}(E) {F_{\nu_{\tau}}(E,X)}
+ N_A\int_E^\infty dE_y F_{\nu_{\tau}}(E_y,X)\frac{d\sigma^{NC}}{\!\!\!\!\!dE}
 (E_y,E)
$$

\begin{figure}[ht]
\centerline{\epsfxsize=3.5in
\epsfbox{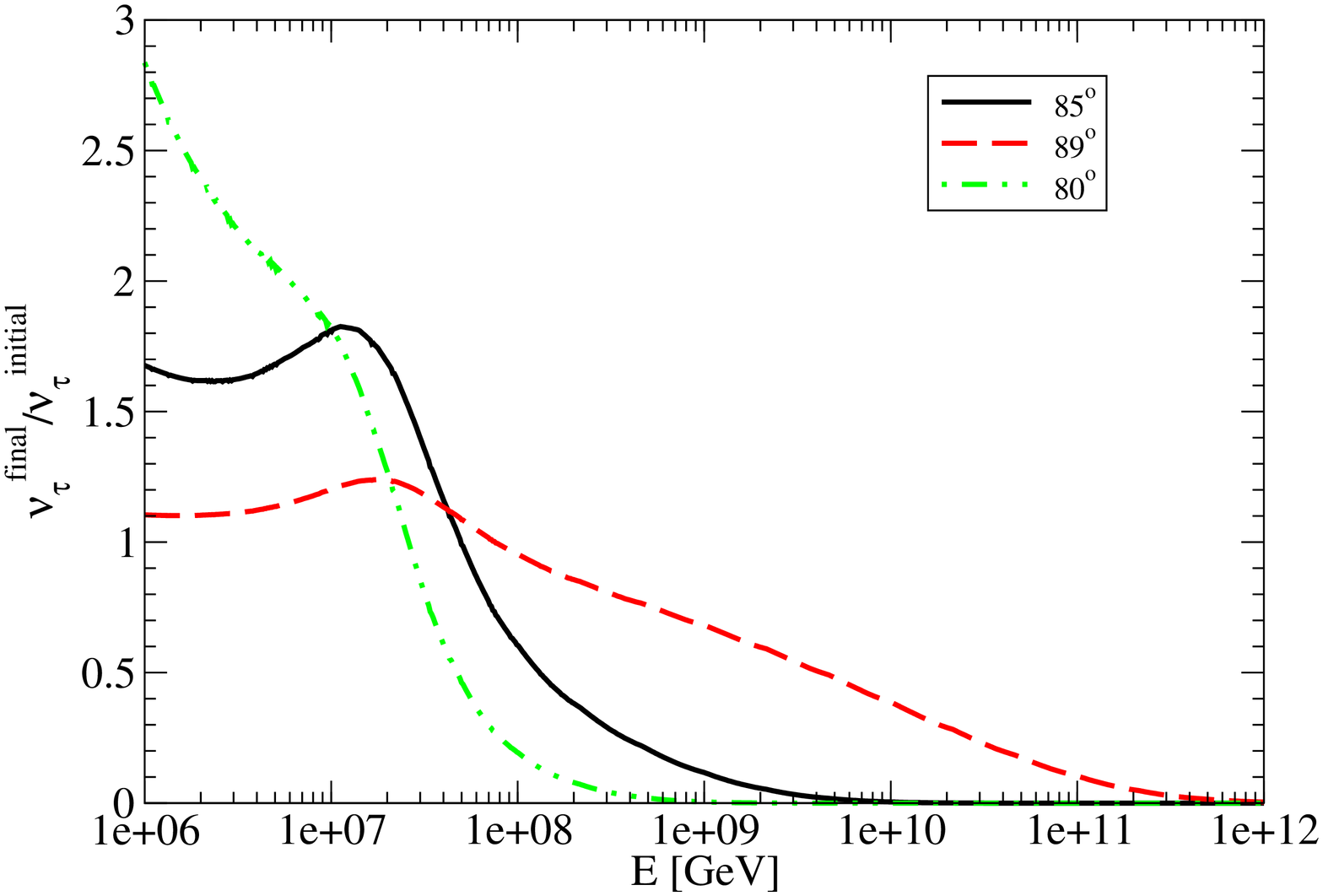}}   
\caption{Ratio $\nu_\tau/\nu_\mu$ for GZK neutrinos, 
at nadir angles of $85^\circ$ and $89^\circ$.}
\label{fig:ratio}
\end{figure}    
$$
+ \int_E^\infty dE_y \frac{F_{\tau}(E,X)}{\lambda_\tau^{dec}}
\frac{dn}{dE}(E_y,E)
\label{nuprop}
$$

\be
 \frac{\partial F_\tau(E,X)}{\partial X}=  
        - \frac{F_\tau(E,X)}{\lambda_\tau^{dec}(E,X,\theta)}
+ N_A 
\int_E^\infty dE_y F_{\nu_{\tau}}(E_y,X)\frac{d\sigma^{CC}}{\!\!\!\!\!dE} 
(E_y,E)
\label{tauprop}
\ee
\be
-\frac{dE_\tau}{dX}=\alpha+\beta E_\tau
\label{eloss}
\ee
Here 
$F_{\nu_{\tau}}(E,X)=dN_{\nu_\tau}/dE$ and $
 F_\tau(E,X)=dN_\tau/dE$ are the differential energy
spectra of tau neutrinos and taus
respectively, for lepton energy $E$, 
at a column depth $X$ in the medium defined by
\be
X = \int_0^L\rho(L')dL'.
\ee

For tau neutrinos, we take into account the attenuation by charged current 
interactions, the shift in energy due to neutral current interactions and the
regeneration from tau decay. For tau leptons we consider their production in 
charged
current $\nu_\tau$ interactions, their decay, as well as 
electromagnetic energy loss.

The effective decay length of produced taus does not go above $10^7$ cm,
even for $E_\tau=10^{12}$ GeV.
This is because electromagnetic energy loss over that distance reduces the
tau energy to about $10^8$ GeV, at which point the tau is more likely 
to decay than interact electromagnetically \cite{10}.

\begin{figure}[t]
\centerline{\epsfxsize=3.5in\epsfbox{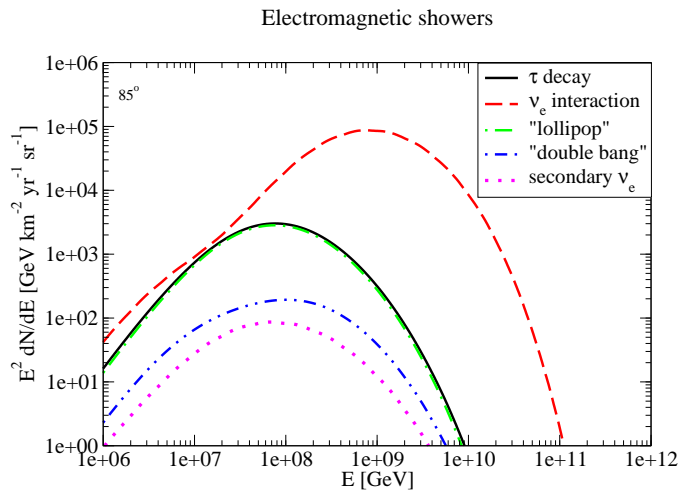}}   
\caption{Electromagnetic showers for GZK neutrinos.}
\label{em}
\end{figure}

We have found that the $\nu_\tau$ 
flux above $10^8$ GeV resembles the $\nu_\mu$ flux. 
The lore that the Earth is transparent to tau neutrinos is not applicable
in the high energy regime. Tau neutrino pileups at small angles with respect to
the horizon are significantly damped due to tau electromagnetic energy loss above
$E_\tau\sim 10^8$ GeV if the column depth is at least as large as the neutrino
interaction length.

At lower energies, $E \leq 
10^8$ GeV, regeneration of $\nu_\tau$ becomes important for trajectories 
where the other flavors of neutrinos are strongly
attenuated, but the $\nu_\tau$ regeneration is very effective. 
The regeneration effect depends strongly on the shape of the initial 
flux and it is larger for flatter fluxes.
The enhancement due to regeneration 
also depends on the amount of 
material traversed by neutrinos and leptons, i.e. on 
nadir 
angle. For GZK neutrinos, we have found that 
the enhancement peaks between $10^6$ and a few$\times 10^7$ GeV depending on
trajectory.


Fig. \ref{fig:ratio} shows the ratio of the tau neutrino flux after
propagation  to incident tau neutrino flux, for $89^\circ$,
$85^\circ$and 
$80^\circ$.
This ratio illustrates a combination of the 
regeneration of $\nu_\tau$ due to tau decay and the attenuation of all 
neutrino fluxes. 
For $89^\circ$, where both the total distance and the density are smaller, 
the attenuation is less dramatic, and the flux can be significant
even at high energy. The regeneration in this case can add about $25\%$ 
corrections at energies between $10^7$ and $10^8$ GeV. 
For $85^\circ$ the relative enhancement is around $80\%$ and peaked at 
slightly lower 
energies, while at $80^\circ$ it is almost a factor of 3 at low energy. At 
$80^\circ$, however, the flux is very strongly attenuated for energies above 
a few $\times 10^7$ GeV. 
It is already clear from here that the total rates will be dominated by the 
nearly horizontal trajectories 
that go through a small amount of matter. However, rates can get significant 
enhancements at low energies where the regeneration from tau decays adds an 
important contribution even for longer trajectories.

\section{Showers}

\begin{figure}[t]
\centerline{\epsfxsize=3.5in\epsfbox{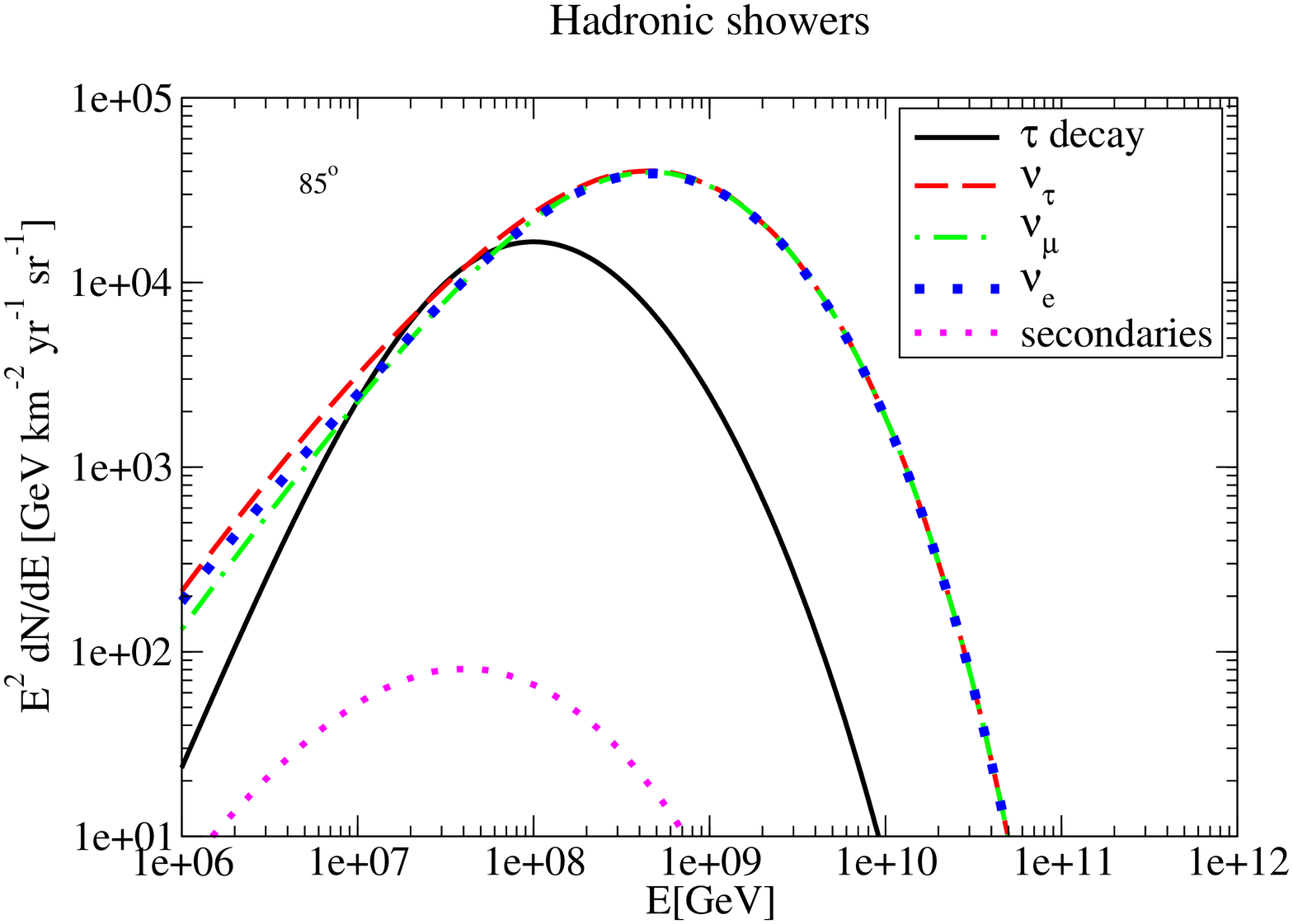}}   
\caption{Hadronic showers for GZK neutrinos.}
\label{had}
\end{figure}

We have translated the neutrino fluxes and tau lepton fluxes into rates for 
electromagnetic and hadronic showers at selected angles to see the effect
of attenuation, regeneration, and the different energy dependences of the
incident fluxes.  We have focused on comparing the $\nu_\tau$ contribution
to the $\nu_e$ and $\nu_\mu$ contributions to determine in what range, if any,
$\nu_\tau$'s enhance shower rates. Electromagnetic shower distributions 
for a nadir angle of $85^\circ$ are shown in Fig. \ref{em}, while 
Fig. \ref{had} shows 
hadronic showers.


\begin{figure}[t]
\centerline{\epsfxsize=3.5in\epsfbox{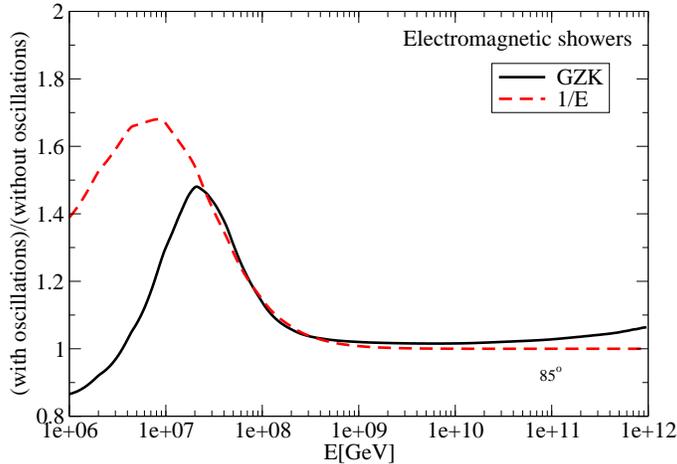}}  
\caption{Ratio of electromagnetic shower rates in the presence
and absence of 
$\nu_\mu\to\nu_\tau$ oscillations for GZK and $1/E$ neutrino spectra 
for nadir angle $85^\circ$ for a km size detector.}
\label{fig:shratioem}
\end{figure}

Fig. \ref{fig:shratioem} shows the ratio of the electromagnetic 
shower rates at nadir angle $85^\circ$ in the presence and absence of 
oscillations for the GZK and $Z$ burst 
neutrino fluxes (which have a characteristic
$1/E$ energy dependence). In absence of oscillations, the only contribution to 
electromagnetic showers comes from $\nu_e$ interactions. In the presence of 
$\nu_\mu\to\nu_\tau$ oscillations, electromagnetic decays of taus from tau 
neutrinos add significant contributions to these rates at energies below 
$10^8$ GeV. In the same time, for the GZK flux, $\nu_e\to\nu_{\mu,\tau}$ 
oscillations
reduce the number of $\nu_e$'s at low energy, such that below a few 
$\times 10^6$ GeV there are fewer electromagnetic showers than in the absence 
of oscillations.

The $\nu_\tau$ flux 
enhancements depend on the shape of the initial flux. The electromagnetic
showers are more sensitive to this shape than hadronic ones. The relative 
enhancement in hadronic showers is also smaller than for the electromagnetic 
showers. This is because for the electromagnetic signal the only contribution 
in the absence of taus is from electron neutrinos, while for hadrons the
tau contribution is compared to a much larger signal, from the interactions
of all flavors of neutrinos.  We have considered contribution from 
secondary neutrinos, which 
we find to be relatively small for all fluxes.  

For kilometer-sized detectors, at for example a nadir angle of $85^\circ$, 
the maximal enhancement due to 
$\nu_\tau$ contribution to electromagnetic shower rates  
for the GZK flux is 
about $50\%$ at $3 \times 10^7$ GeV, while for a $1/E$ flux, it is even 
larger, 
about $70\%$, 
at slightly lower energy. These energy ranges are relevant for IceCube.  
 For energies relevant to RICE,
tau neutrinos do not offer any appreciable
gain in electromagnetic shower signals compared to $\nu_e\rightarrow e$ CC
interactions, and they contribute at essentially the same level as $\nu_\mu$
to hadronic shower rates through NC interactions.

\begin{figure}[t]
\centerline{\epsfxsize=3.5in\epsfbox{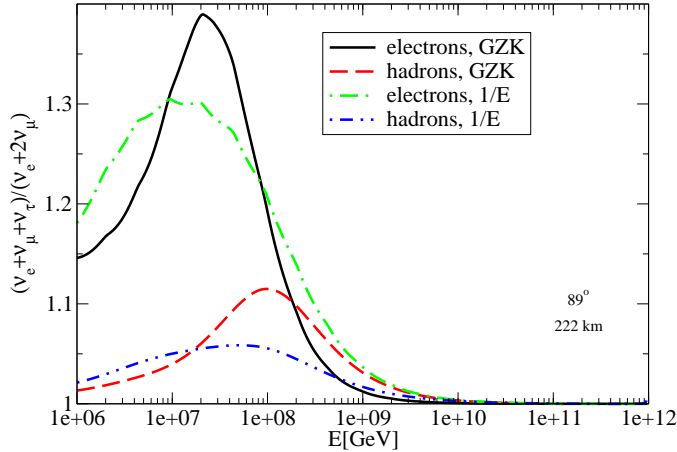}}   
\caption{Ratio of electromagnetic and hadronic shower rates in the presence
and absence of $\nu_\mu\to\nu_\tau$ oscillations for GZK  and
$1/E$ fluxes of neutrinos
 for detection over 222 km of ice.}
\label{shratioanita}
\end{figure}

In 
Fig. \ref{shratioanita}
we show 
the ratio of electromagnetic and hadronic shower
rates in the presence and absence of $\nu_\mu\to\nu_\tau$ oscillations.  
 The maximum
enhancement due to the presence of $\nu_\tau$ is about 40\% at this angle, as
expected since this trajectory has low column density. However, as previously
discussed, the enhancement occurs in a larger energy range and it peaks at
higher energy than in the case of a small size detectors. In 1 km only taus
with energies below a few $\times 10^7$ GeV have a significant probability
to decay, while much higher energy taus can decay over the total distance of
more than 200 km.

At very high energies tau neutrinos do not contribute large signals
to kilometer-sized detectors because high energy tau 
decay lengths 
are very large, so the probability of a tau decaying in the detector is low.
For detectors like ANITA which can sample long trajectories through the
ice, one would expect a larger tau neutrino contribution to the signal from 
tau decay.
Despite the long trajectory (222 km with a maximum depth of 1 km for a
neutrino incident at $89^\circ$ nadir angle) the tau contributions to
the electromagnetic shower rate is quite small for fluxes expected to 
contribute
in the ANITA signal. 
For hadronic showers, the suppression of $\tau$ decay 
to hadrons
relative to $\nu_e$ NC interaction contributions is about the same
as for electromagnetic showers compared to $\nu_e\rightarrow e$. The $\nu_\tau$ contribution to the
hadronic shower rate from interactions is about the same as the $\nu_e$ contribution.
Clearly, large detectors with energy threshold below $10^8$ GeV
and with very good angular resolution
are needed for distinguishing between diffrerent neutrino
flavors.

Recently, it has been noted that rock salt formations have
similar properties to the Antarctic ice and can therefore be used as
large scale neutrino detectors \cite{salsa}. Salt has a higher density
($\rho_{salt}=2.2$ g/cm$^3$) than ice ($\rho_{ice}=0.9$ g/cm$^3$),
so it is possible to achieve an effective detection volume of several
hundred km$^3$ water equivalent in salt. This is somewhat better
than RICE but with a much smaller actual detector size.
The threshold for detecting the radio signal from showers in
salt is of the order of $\sim 10^{7}$ GeV.  Another recenly proposed 
experiment is 
 is LOFAR \cite{lofar}, a digital telescope array
designed to detect radio Cherenkov emission in air showers. LOFAR has
sensitivity in an 
energy range of $\sim 10^{5} - 10^{11}$ GeV, so it can detect showers
at much lower energies than other radio Cherenkov experiments.
LOFAR will likely be configured to detect horizontal showers from skimming
neutrinos as well. With its low energy threshold, LOFAR has an excellent
opportunity to observe the shower enhancement at lower energies due to
$\nu_{\tau}$ regeneration and tau pileup, which is not easily
accessible in ANITA.

\section*{Acknowledgments}

This work was supported in part by the Department of Energy under
contracts DE-FG02-91ER40664, DE-FG02-95ER40906, DE-FG02-93ER40792, 
DE-FG02-04ER41298 and DE-FG02-04ER41319.  


\end{document}